\documentclass[aps,prd,twocolumn,superscriptaddress,showpacs]{revtex4}
\usepackage{graphicx}
\usepackage{dcolumn}
\usepackage{bm}
\usepackage{natbib}
\usepackage{multirow}
\newcommand{\beq}{\begin{equation}}
\newcommand{\eeq}{\end{equation}}
\newcommand{\beqa}{\begin{eqnarray}}
\newcommand{\eeqa}{\end{eqnarray}}

\def\affilmrk#1{$^{#1}$}
\def\affilmk#1#2{$^{#1}$#2;}

\def\zrh{1}
\def\ar{2}

\begin{document}

\title{Impact of Dark Matter Microhalos on Signatures for Direct and Indirect Detection}
\author{
Aurel Schneider\affilmrk{\zrh}, Lawrence Krauss\affilmrk{\zrh,\ar} and Ben Moore 
}
\address{
{Institute for Theoretical Physics, University of Zurich, Zurich, Switzerland}
\affilmk{\ar}{School of Earth and Space Exploration and Department of Physics, Arizona State University, PO Box 871404, Tempe, AZ 85287}
}

\date{\today}

\begin{abstract}

Detecting dark matter as it streams through detectors on Earth relies on knowledge of its phase space density on a scale comparable to the size of our solar system. Numerical simulations predict that our Galactic halo contains an enormous hierarchy of substructures, streams and caustics, the remnants of the merging hierarchy that began with tiny Earth mass microhalos. If these bound or coherent structures persist until the present time, they could dramatically alter signatures for the detection of weakly interacting elementary particle dark matter (WIMP). Using numerical simulations that follow the coarse grained tidal disruption within the Galactic potential and fine grained heating from stellar encounters, we find that microhalos, streams and caustics have a negligible likelihood of impacting direct detection signatures implying that dark matter constraints derived using simple smooth halo models are relatively robust. We also find that many dense central cusps survive, yielding a small enhancement in the signal for indirect detection experiments.

\end{abstract}

\maketitle

\setcounter{footnote}{0}

\section*{Introduction}

In a $\Lambda$CDM dominated universe, structure forms by the hierarchical clustering and merging of small density perturbations \cite{Peebles1982}. 
Numerical simulations that follow these processes predict that our Galactic halo should contain a vast hierarchy of surviving substructures - the remnants of the entire halo merger tree \cite{Moore1999}. The number density of substructures of a given mass $M$ goes as $n\propto M_{subs}^{-1}$ and they span over 15 decades in mass \cite{Ghigna1998}. The smallest, oldest and most abundant are Earth-mass microhalos with a half mass radius of $10^{-2}$ pc that formed at $z\simeq80-30$ \cite{Green2004,Bergstrom1999,Berezinsky2003,Diemand2005,Koushiappas2009}. This minimum mass is modulated by the free streaming velocity which is related to the mass of the neutralino \cite{Hofmann2001,Bertschinger2006}.

Simulations of relatively large subhalos suggest that their gravitational interactions with a disk potential, can lead to a destruction of subhalos at distances closer than 30 kpc \cite{dOnghia2010}. Smaller subhalos form earlier, however, with denser cores, and are therefore the most probable dark matter structures to survive gravitational interactions. A second source of fine grained structure to survive are the numerous caustic sheets and folds that form due to the very high initial phase space density of the cold dark matter particles \cite{Sikivie1999}. These are wrapped in a complex way within all the subsequent structures that form, however in the absence of a heating term, the fine grained phase density would be preserved.

With low internal velocity dispersion and high mean density, both the event rate and the characteristic spectrum of energy deposited by dark matter in direct detection experiments \cite{Copi2007} could be affected by any features surviving in the phase space distribution of CDM particles. Direct detection experiments are sensitive to the density and velocity distributions of WIMPs on a scale of $\approx 10^{13}m$, the distance the Earth travels over a year. In order to make predictions and exclusion limits, these experiments assume that the dark matter is completely smooth on these scales, with a well mixed Maxwell-Boltzmann velocity distribution \cite{CDMS2009,DAMA2010}. Furthermore, if any such small high-density clumps actually dominate the dark matter distribution in the solar neighborhood, the indirect detection signal due to dark matter annihilation in the galaxy might also be affected. 

Since existing N-body simulations of galaxy formation do not have a resolution that goes down to objects with mass as small as $10^{-6} M_{\odot}$, in order to address the question of the survival and impact of such microhalos we need to combine analytical estimates with the results of smaller scale simulations that can resolve such objects.

Previous studies have made analytic \cite{Zhao2007} and numerical estimates \cite{Goerdt2006} of the disruption timescale of microhaloes as they orbit through the stellar field. Zhao et. al. \cite{Zhao2007} argued that most of the microhaloes should be completely destroyed by encounters with stars, whilst Goerdt et. al. \cite{Goerdt2006} show that indeed, whilst most of the mass is unbound the dense central cusp may survive intact.

We extend this work by numerical calculating the tidal disruption of microhaloes as they actually orbit through a field of stars, as well as self-consistently including the Galactic halo potential. We calculate the survival statistics of microhaloes using realistic orbital distributions within the disk, allowing us to follow the dynamical structure of the dark matter streams and thus to estimate the fine grained phase space distribution function of WIMPs on scales relevant to dark matter detection experiments.

\section*{Microhalo parameters and disruption processes}

The dominant processes that can affect microhalos involve gravitational interactions with baryons in the stellar field during the crossing of the disk and also tidal effects of the disk potential during the orbit of the microhalo. Unfortunately we cannot account for both effects simultaneously in our simulations as it would require setting up a self-consistent disk with billions of stars. Instead we look separately at the effects of stellar disruption and tidal streaming and we then estimate the combined behavior.

As a substructure halo crosses through a stellar field, high-speed interactions with single stars will heat up the halo distribution, causing it to increase its velocity dispersion and hence its scale size will grow. This process is analogous to galaxy harassment that occurs in clusters \cite{Moore1996} and basically has a timescale proportional to the relaxation time of the stellar disk and the time each microhalo spends within the fluctuating potential field of the stars. For an analytical estimation of this process we follow the Goerdt et. al. paper \cite{Goerdt2006}.

In the 'distant-tide' approximation \cite{B&T} the internal energy increase of the microhalo due to a single encounter with a fixed star is given by
\begin{equation}\label{heating}
\delta E(b) = \frac{1}{2}\left(\frac{2GM_{*}}{b^{2}V_{mh}}\right)^2\frac{2}{3}\langle r^2\rangle,
\end{equation}
where $b$ is the impact parameter and $V_{mh}$ is the velocity of the microhalo. Since halos with an early formation time have a low concentration we can set $\langle r\rangle\approx 0.5r_{vir}$.

One encounter can totally disrupt the microhalo if $\delta E$ exceeds the binding energy $E_b$. Since $E_b\approx 0.4v_{vir}^2$ with $v_{vir}^2=Gm_{vir}/r_{vir}$ \cite{Spitzer1987}, the minimal encounter parameter that does not entirely disrupt the microhalo is found to be
\begin{equation}
b_{min}\approx 0.8\left(\frac{GM_{*}r_{vir}}{V_{mh}v_{vir}}\right)^{1/2}.
\end{equation}
We can now define the disruption probability of a microhalo in a stellar field
\begin{equation}\label{sigma}
p=\frac{1}{E_b}\int \delta E dN = \int_{0}^{b_{min}}dN + \frac{1}{E_b}\int_{b_{min}}^{\infty} \delta E dN,
\end{equation}
where $dN=2\pi n b db V_{mh} dt$. Here we have used $\delta E=E_b$ for $b<b_{min}$. Performing the integration leads to a disruption probability of
\begin{equation}
p\approx 4GM_{*}n_*t\left(\frac{r_{vir}}{v_{vir}}\right)=\frac{2GM_*n_*t}{5H_0\Omega_{m0}^{1/2}}(1+z)^{-3/2},
\end{equation}
where we have used the definition of the virial radius $M_{vir}=\frac{4\pi}{3}r_{vir}^3200\rho_c$ with $\rho_c=\frac{3H^2}{8\pi G}$. The microhalo is completely destroyed at $p=1$. Therefore we get the average disruption time
\begin{equation}\label{disruption_time}
t=250\left(\frac{0.04M_{\odot}pc^{-3}}{M_*n_*}\right)\left(\frac{1+z}{61}\right)^{3/2}Myr. 
\end{equation}
A microhalo with a formation redshift $z\sim60$ should therefore survive about $250$ Myr in a stellar field with a density similar to the one in the solar neighbourhod. However this is only true on average, since one very close encounter can immediately lead to total disruption.

The above estimate does not take into account the internal structure of the microhalos and should therefore only give a very rough estimation of the disruption time. Also it does not allow one to follow the mass loss during the disruption process. We therefore perform a simulation where a microhalo is crossing a periodic box of stars. The box has a length of $50$ pc and is filled with randomly distributed stars with the density $\rho=0.04$M$_{\odot} $pc$^{-3}$ and the velocity dispersion $\sigma = 50$ km/s. This constellation corresponds to the stellar field in the disk at the solar radius \cite{B&T}. For simplicity all the stars have the average mass of $0.7$ M$_{\odot}$. The microhalo which is crossing the box at $200$ km/s is set up by the halogen-code of Zemp et. al. \cite{Zemp2008}. Corresponding to the results in \cite{Diemand2005} it has a mass of $10^{-6}$M$_{\odot}$ and the density profile
\begin{equation}
\rho(r)\propto \frac{1}{\left(\frac{r}{r_s}\right)^{\gamma}\left(1+\left(\frac{r}{r_s}\right)^{\alpha}\right)^{\frac{\beta-\gamma}{\alpha}}}
\end{equation}
with $\alpha=1$, $\beta=3$ and $\gamma=1.2$, as well as a concentration parameter of $c=r_{200}/r_s=1.6$. The virial radius $r_{200}$ is defined with respect to the background density at $z=60$, the average formation redshift of a microhalo \cite{Green2004}.

\begin{figure}[ht]
\centering
\includegraphics[scale=.42]{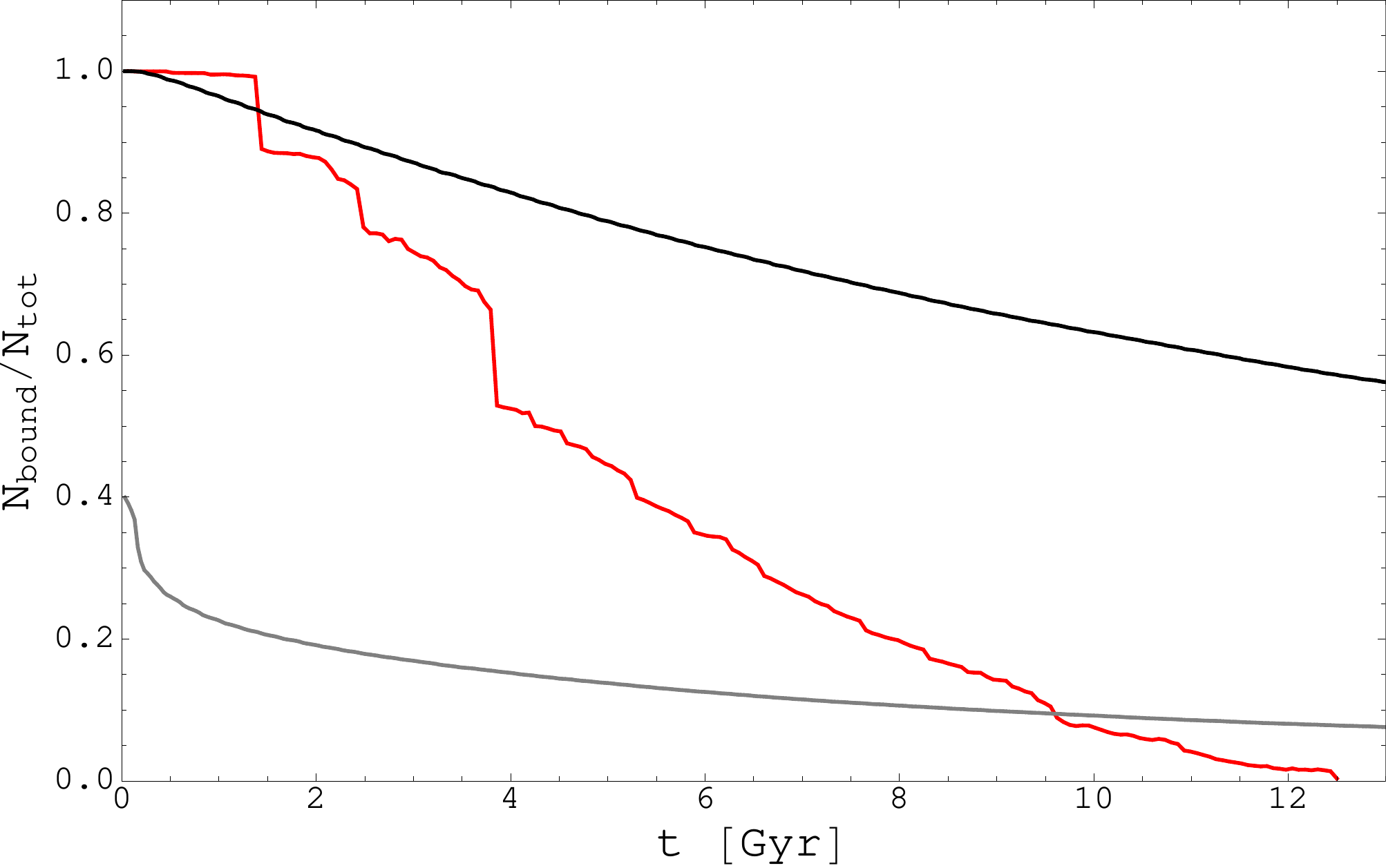}
\caption{\textit{Ratio between bound and total mass of a microhalo crossing a stellar field (red) and orbiting in a Milky Way potential (black, grey). The red curve is stretched in order to simulate the effect of disk crossing on an average microhalo with about 80 disk crossings. The black and the grey curve correspond to a microhalo that spent 0 Myr respectively 80 Myr in the stellar field before orbiting around the galactic potential. The dominant mass loss is coming from stellar interaction until complete disruption after about 12 Gyr.}}
\label{disruption_plot}
\end{figure}

In our simulation we find that $50\%$ of the microhalo mass is unbound after 80 Myr of box-crossing (which corresponds to about 40 perpendicular disk passages). After 160 Myr (80 disk passages) even the central core starts to disappear and more than $90\%$ of the microhalo is completely disrupted (see pictures in Table \ref{mh_crossing_den}). At latest after 200 Myr (100 disk passsages) no bound structure is left (see Fig \ref{disruption_plot}). Our simulation gives therefore a slightly shorter disruption time than the simplified analytical estimation of equation (\ref{disruption_time}).

\begin{table}[ht]
\centering
\begin{tabular}{ccc}
\includegraphics[scale=0.097]{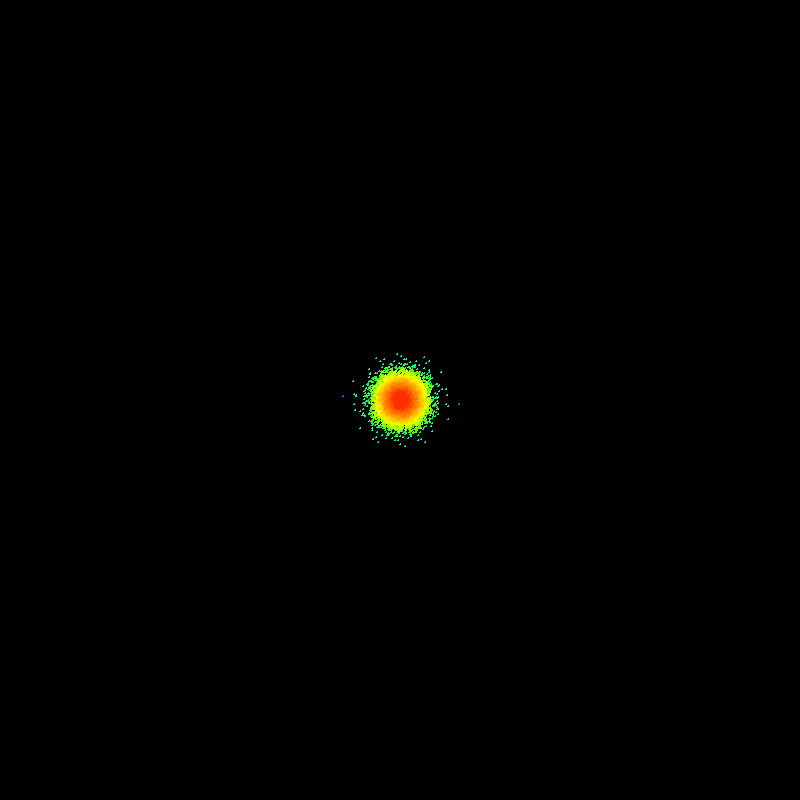}&\includegraphics[scale=0.097]{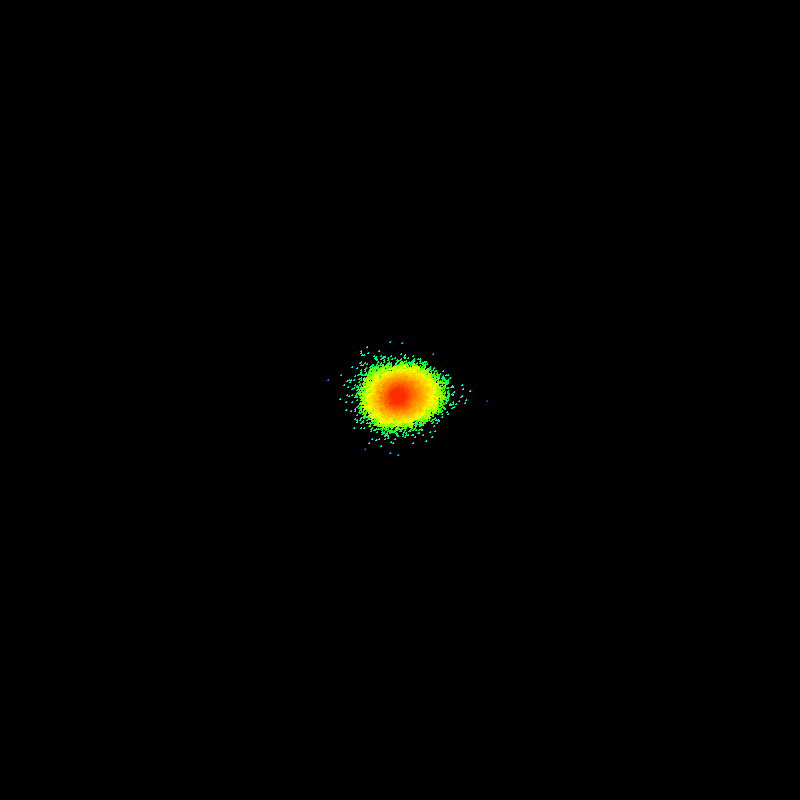}&\includegraphics[scale=0.097]{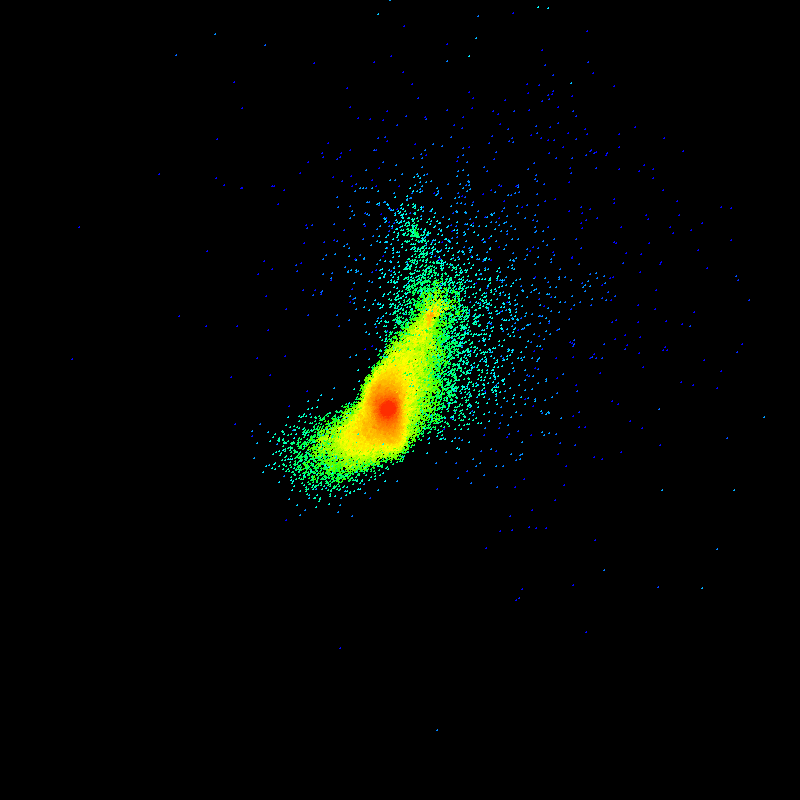}\\
\newline
\includegraphics[scale=0.097]{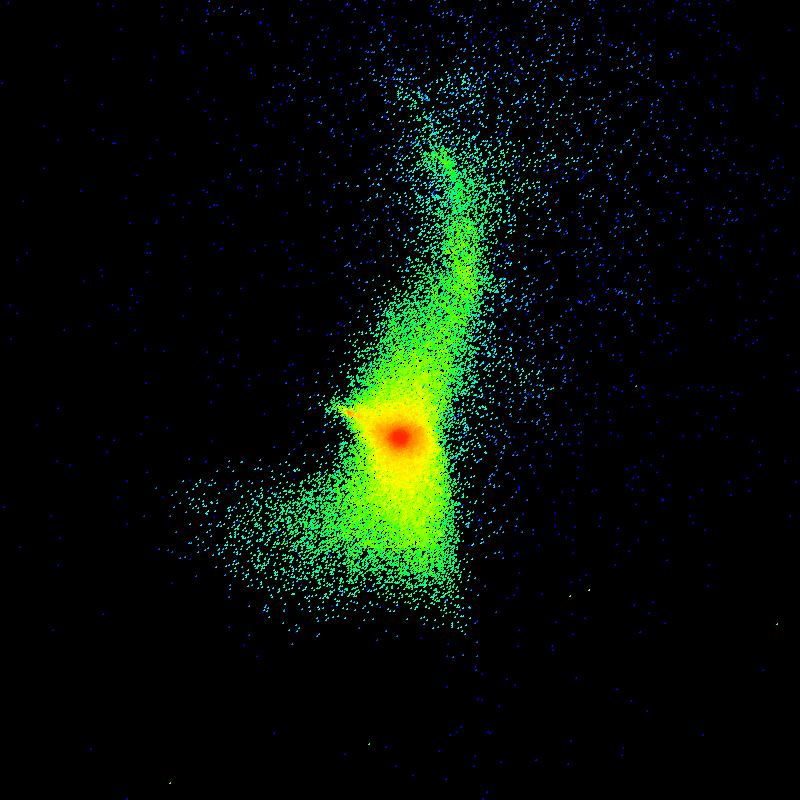}&\includegraphics[scale=0.097]{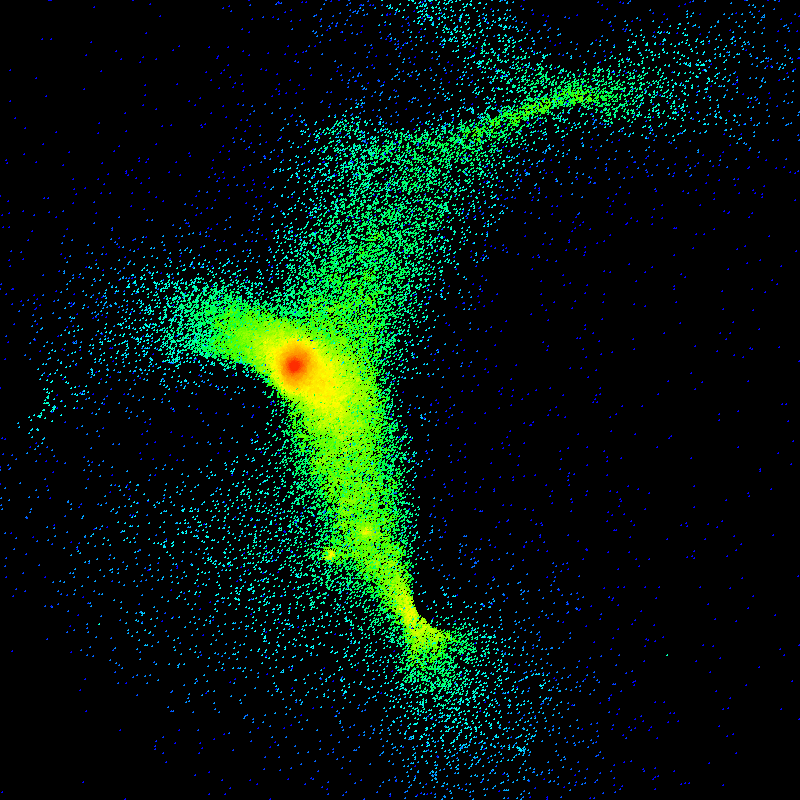}&\includegraphics[scale=0.097]{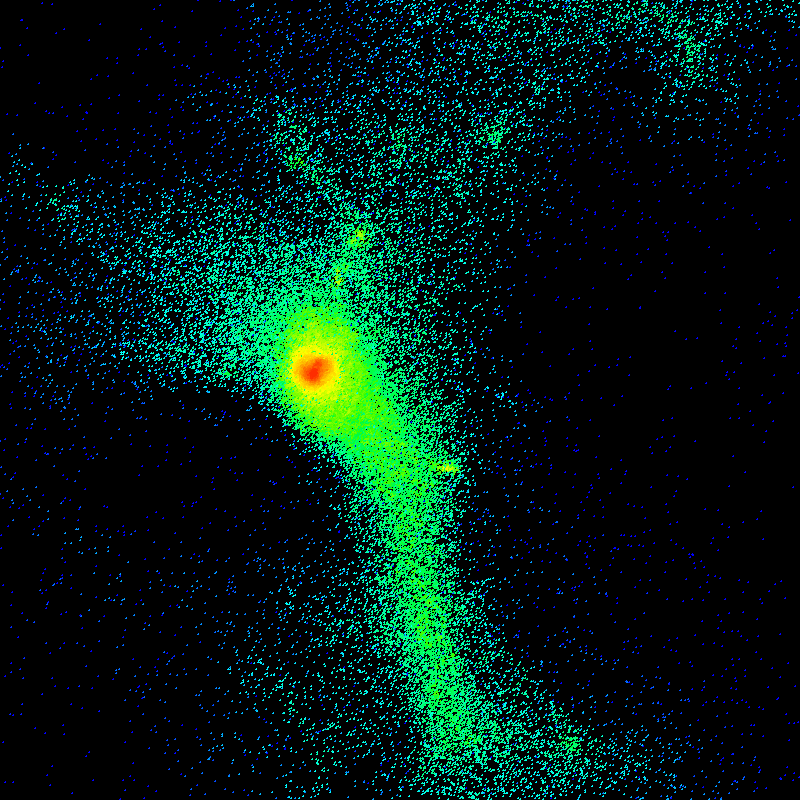}\\
\newline
\includegraphics[scale=0.097]{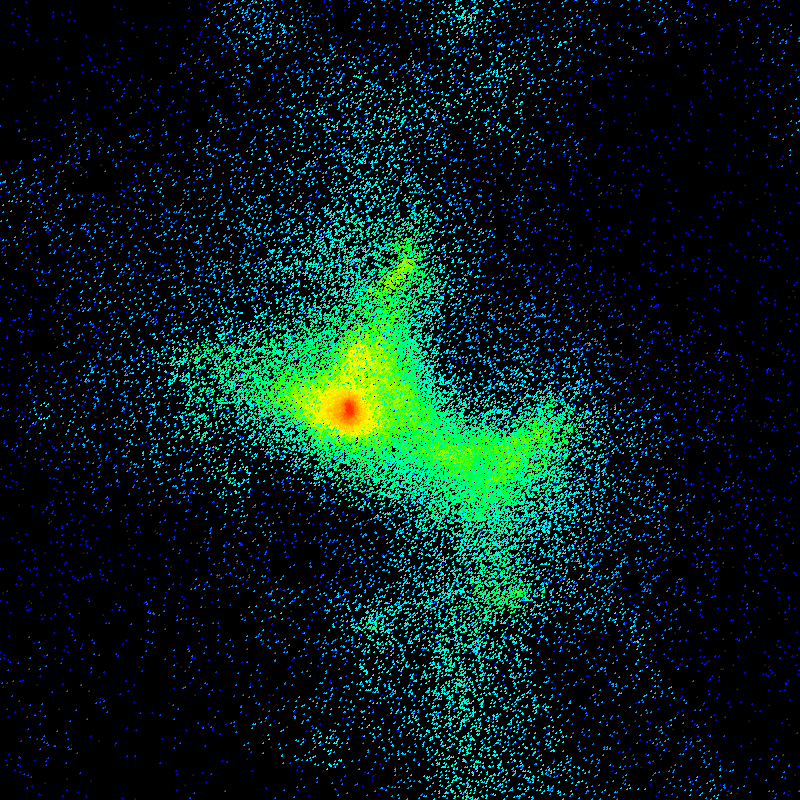}&\includegraphics[scale=0.097]{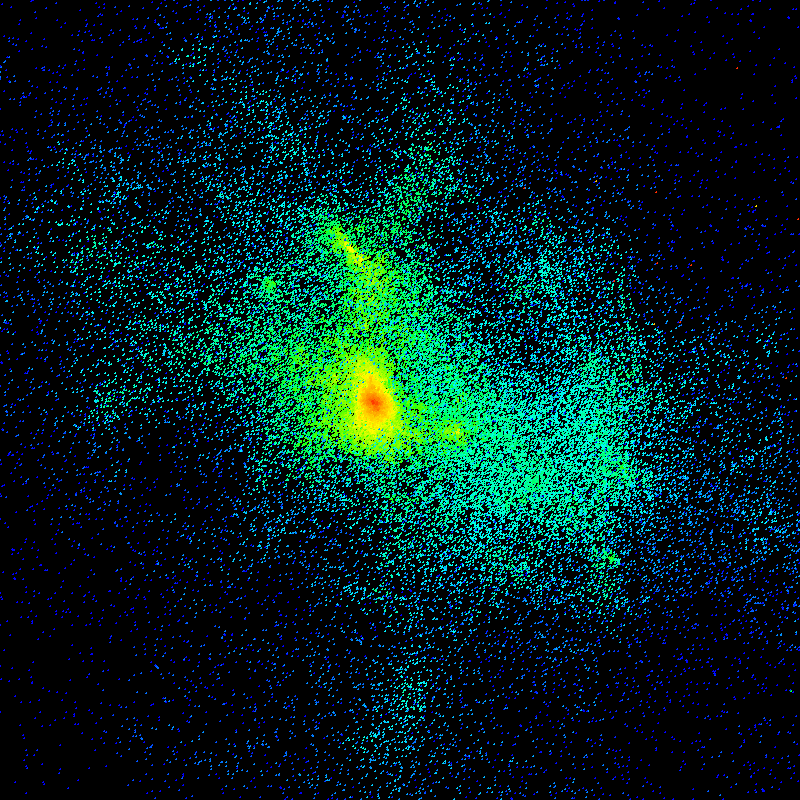}&\includegraphics[scale=0.097]{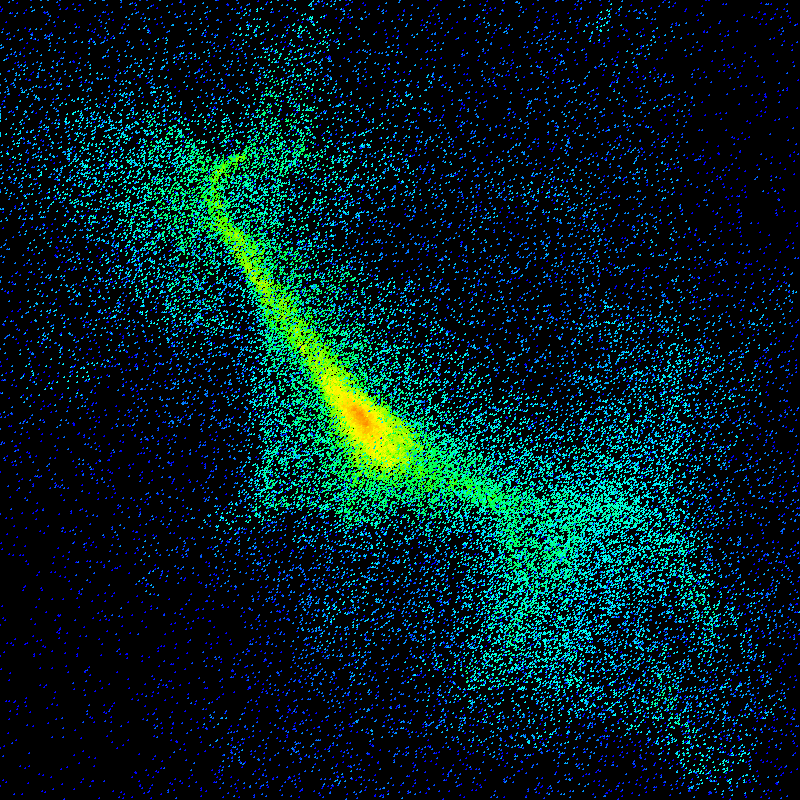}\\
\end{tabular}
\caption{\textit{Microhalo density map at $t=$ 0, 20, 40, 60, 80, 100, 120, 140, 160 Myr (from the upper left to the lower right). The boxlength of the images is 0.38 pc.}}
\label{mh_crossing_den}
\end{table}

In order to determine what fraction of microhalos survive until the present day, we have to calculate orbital statistics and the distribution of disk crossing times. This can be established by tracing back the orbits of particles in a galactic potential. We use the standard Milky Way model with disk and halo particles set up by the GalactICS code \cite{Widrow2005} and we select a sample of halo particles in a small box around the position of the sun. The orbits of these particles are followed backwards in time and we find that the average number of disk crossings for these particles is $\overline{c}=80$ with a standard deviation of $\sigma_{c} =43$. The average crossing radius is (not surprisingly) $\overline{R}=8$ kpc with $\sigma_{R} =4$ kpc. The spread of disk crossing events for different particles follows a Maxwell-Boltzmann distribution.

We use this disk crossing distribution combined with the rate of mass loss determined from our numerical study to calculate the survival statistics of microhalos in the vicinity of the sun. Since the timescale for complete disruption in our simulation is equivalent to the average time a microhalo spends in the stellar disk, we conclude that the average microhalo in the vicinity of the sun is just about to be entirely destroyed at the present time (see also \cite{Green2007}). At most five percent of its initial mass is still in a bound core. However the spread in the number of disc crossings is relatively wide and a significant fraction of microhalos should still have surviving cores. Mass loss is nevertheless important: microhalos maintaining more than $50\%$ of their initial mass should be rare. Figure \ref{disruption_plot} illustrates the mass loss, where the red curve shows the disruption of a typical microhalo with 80 disk crossings in 10 Gyr at the radius of the sun.

However, disk crossing is not the only source of dynamical disruption. While orbiting the galaxy, a microhalo is under the constant influence of the global Galactic potential, and tidal forces will act so that the microhalo's structure becomes elongated and unbound particles will form leading and trailing tidal streams. The detailed impact of tidal streaming depends on the orbit of the microhalo and on the shape of the host potential. In our simulations we use a disk potential that emerges from a density distribution of the form
\begin{equation}
\rho(R,z)\propto exp(-R/R_d)sech^2(z/z_d).
\end{equation}
Here $R$ and $z$ are the disk radius and the height respectively, which we set to be $R_d=2.8$ kpc, $z_d=0.4$ kpc. The disk mass is $M_d=4.5\cdot10^{10}M_{\odot}$. In all our simulations the orbit of the microhalo is chosen to be roughly spherical with a distance of $7.9$ kpc from the galactic center.

We cannot model both heating due to stellar interactions and tidal elongation at the same time since this would require following the motion of 50 billion disk stars. We therefore performed orbital simulations for three different cases: an initially completely undisturbed microhalo, a microhalo that first crossed the stellar field for 80 Myr and has lost about 60 percent of its mass, and a completely disrupted microhalo that spent more than 160 Myr in the stellar field.

The length of the tidal streams $l$ due to the orbiting process can be crudely estimated with the relation $l(t) \sim \sigma_{mh} t$, where $\sigma_{mh}$ is the velocity dispersion of the initial microhalo. For the initially unperturbed microhalo $\sigma_{mh}\sim10^{-3}$ km/s, causing a stream length of roughly $l \sim 10$ pc after one Hubble time. For the initially completely disrupted microhalo $\sigma_{mh}\sim10^{-2}$ km/s, and the stream length is about $l \sim 100$ pc after a Hubble time. These length scales agree well with our simulations (i.e. see Table \ref{mh_orbiting}).

\begin{table}[ht]
\centering
\begin{tabular}{c}
\includegraphics[scale=0.3]{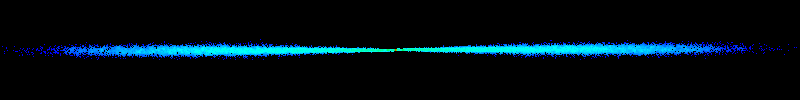}\\
\newline
\includegraphics[scale=0.3]{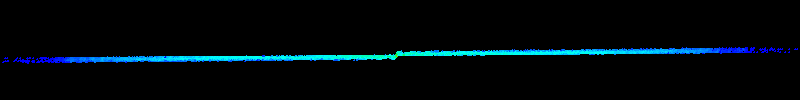}\\
\newline
\includegraphics[scale=0.3]{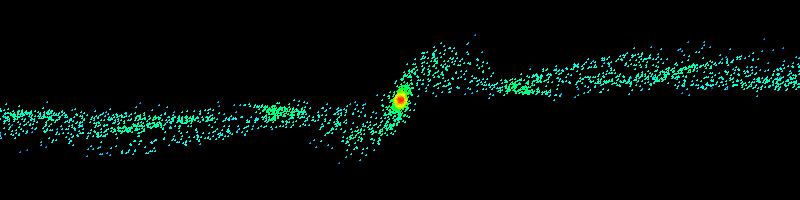}\\
\newline
\end{tabular}
\caption{\textit{Streaming microhalo after 10 Gyr on a roughly circular orbit around a Milky Way potential: The two images on the top show the sheet-like streams from the top and from the side (boxlength: 30 pc). The third image is a zoom in at the centre where the still bound core is visible (boxlength: 2 pc). For these pictures we have used an initially non disrupted microhalo with formation redshift 60.}}
\label{mh_orbiting}
\end{table}

Orbiting in the galactic potential significantly reduces the mass of the microhalo (see black and grey lines in Fig \ref{disruption_plot}). However, the rate of tidal mass loss is suppressed as the tidal radius is steadily reduced. The central cusp of each dark matter microhalo has a very deep potential, as a consequence there is always a bound core remaining, even for a microhalo that has been heated in the stellar field before orbiting.

Comparing the curves in Fig \ref{disruption_plot} leads to the conclusion that disk crossing is the dominant disruption process and the only one that can lead to complete distruction of the microhalo. The step-like decrease of the curve is an indication of very close encounters that play a mayor role in the disruption process. Tidal stripping on the other hand can also significantly reduce the mass but it never completely distroys the microhalo because of its tightly bound inner core.

\section*{Implications for Dark Matter Detection}

In direct detection experiments the differential interaction rate is sensitive to the fine grained density and the velocity distribution of dark matter particles on A.U. scales \cite{Jungman1996,Copi1999}.  Substructures like microhalos can affect the interaction rate if they are abundant enough to have a substantial likelihood of existing in the solar neighbourhood and if their density is at least the same order of magnitude as the background dark matter density in this region,
$ \rho_{bg}\sim 10^{7} M_{\odot}$kpc$^{-3}$ (see for example \cite{Catena2009}).
Equally important, the phase space for the energy deposits associated with dark matter events will not be that appropriate for an isothermal halo if a single microhalo were to dominate the density distribution in the solar neighbourhood \cite{Copi2001,Copi2007}.

\begin{figure}[ht]
\centering
\includegraphics[scale=.44]{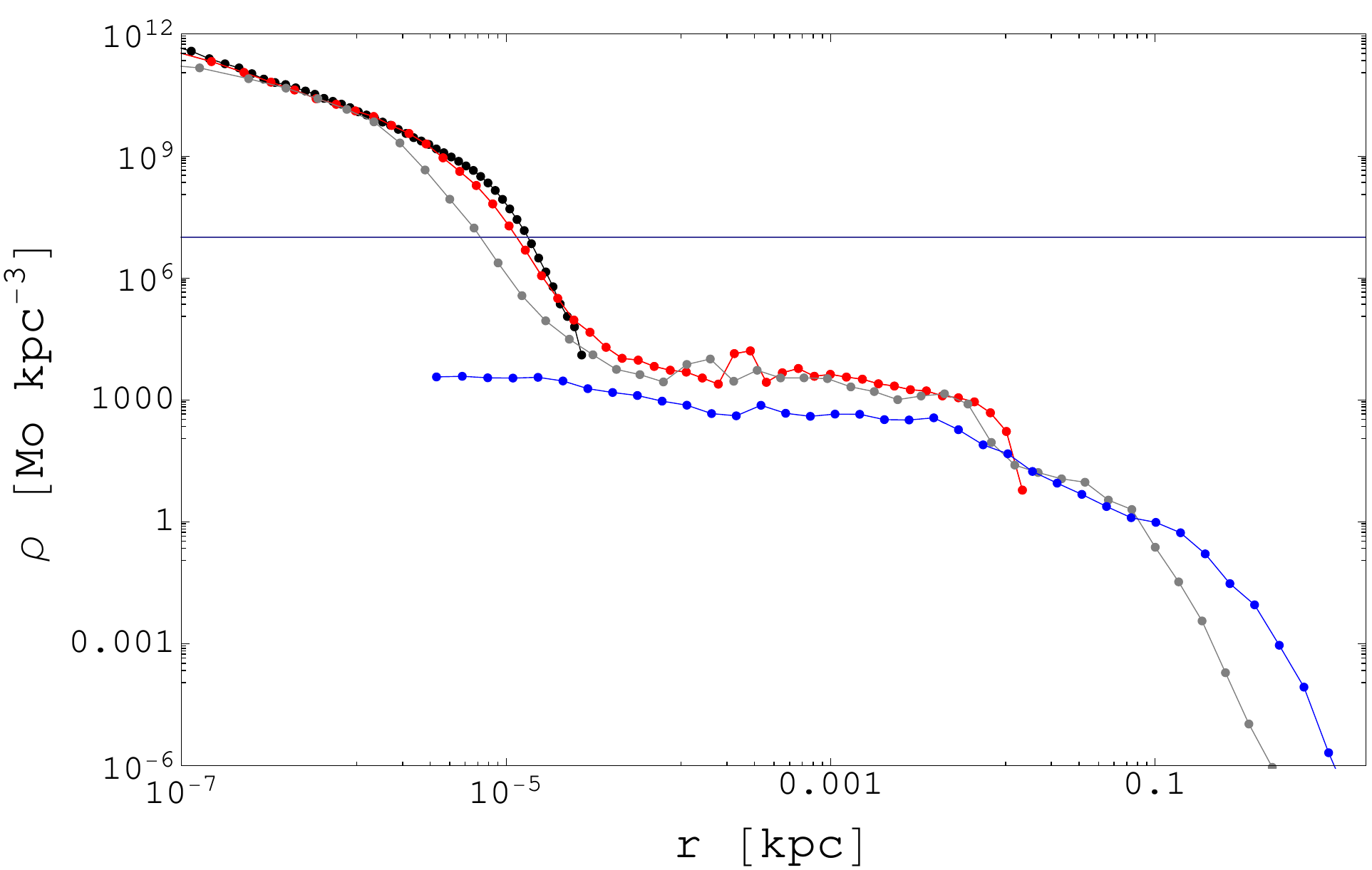}
\caption{\textit{Stream densities of microhalos after an orbital time of ten Gyr. Before orbiting the microhalos have spent a time of 0 Myr (red), 80 Myr (grey) and 160 Myr (blue) in the stellar field. There is still a visible bound core in the red and the grey profile but no more in the blue one. The black dots represent the density profile of a completely undisrupted microhalo that spend no time on an orbit. The straight blue line corresponds to the average dark matter density around the sun.}}
\label{stream_density}
\end{figure}

Our results above suggest that none of these conditions are generally achieved. In Figure \ref{stream_density} we plot the stream densities of microhalos that crossed the stellar field for 0 Myr (red), 80 Myr (grey) and 160 Myr (blue), before orbiting in the galactic potential for 10 Gyr. The tidal streams of the initially unperturbed halo (red) have an average density of $\rho\sim 10^4 M_{\odot}$kpc$^{-3}$, which is already negligibly low compared to the background. Only the very tiny core still maintains its initial density of $\rho \approx 10^{11} M_{\odot}$kpc$^{-3}$. The initially disrupted microhalo (blue) has no more bound core. Its stream density is only at about $\rho\sim 10^2-10^3 M_{\odot}$kpc$^{-3}$. The approach of first measuring the stellar disruption and then looking at tidal effects underestimates the stream density, since the microhalos get most of their heat energy right at the beginning. Therefore, the actual stream density of an average microhalo should be somewhere between the red and the blue line in Figure \ref{stream_density}.

Since the stream densities are far below the value of the local galactic density, only a surviving core existing in the region of the earth would any effect upon direct detection. However, only about half of the microhalos still have bound cores because of disk crossing, and tidal effects further reduce the mass of the cores to less than ten percent of their original value. We note that any substructures orbiting primarily within the disk plane would be quickly destroyed by stellar encounters.

The chance of being in such an overdense region can then be optimistically estimated: An extrapolation of the subhalo mass function leads to a microhalo number density $n_{mh}$ of about 500 pc$^{-3}$ at the solar radius \cite{Diemand2005}. This number can be divided by two due to the disruption processes stated above and again by two since microhalos orbiting in the disk plane are completely disrupted. We then end up with the approximation  of $n_{mh}\sim 100$ pc$^{-3}$ at the solar radius. Each microhalo has a volume of about $V_{mh}\sim 10^{-9}$ pc$^3$ and therefore there is a chance of about $0.0001\%$ of being in such an overdense region.

The streams of particles stripped from microhalos are coherent and long, thus it is appropriate to calculate their volume filling factor. Since the stream density is $\rho\sim 10^2-10^4 M_{\odot}$kpc$^{-3}$ we expect that our solar system is criss-crossed with $f_b \times (10^3-10^5)$ streams, where $f_b\approx0.1$ is the fraction of the local Galactic halo density that forms from substructures up to a solar mass. Larger substructures may be completely disrupted at the Sun's position in the Galaxy due to global disk shocking and tides \cite{dOnghia2010}. The velocity dispersion within an average stream due to heating by disk stars is $\sigma\sim10^{-2}$ km/s. Thus, the local density is determined by the superposition of a large number of independent streams, and the overall velocity distribution at the solar radius should be essentially Maxwellian, isotropic and smooth with no spiky structure, as we would assume for a smooth halo model with no substructures. The signatures of streams could be only be detected experimentally with over several hundred events.

The case for indirect detection is somewhat different from that described above. In indirect detection experiments one tries to detect the annihilation products, such as gamma-rays, coming from the highest density dark matter regions, which is proportional to the square of the dark matter density times the volume of the region observed \cite{Lake1990,Calcaneo2000,Fermi2010,Bertone2005}.  Consider a volume containing on average one microhalo $V \approx 10^{-2}$ pc$^{-3}$. The luminosity due to the smooth background is therefore $L_{bg} \propto V \rho^2 = 10^{-6} M_{\odot}^2 $pc$^{-3} $, whereas the luminosity of a surviving microhalo core is 
\begin{equation} 
L_{mh} \propto V_{core} \rho_{core}^2 \approx 5 \times 10^{-7} M_{\odot}^2 pc^{-3},
\end{equation}
where we have assumed a mean core density of $10^{10} M_{\odot} $pc$^{-3}$. Thus the net boost factor due to microhalos is about 1.5, and stays below the detection limits of the FERMI experiment \cite{Pieri2005}. However this number is highly uncertain since it depends on extrapolations of both the substructure mass function and also on the microhalo internal density structure.

Finally, we consider the evolution of caustic sheets of particles within the Galactic halo. In the absence of fine grained heating, narrow sheets and folds will occupy regions of phase space within all collapsing CDM structures. As structures merge hierarchically, these caustic features become wrapped in phase space like a fine fabric that has been crumpled into a ball, the phase space density at any point being preserved.
During the matter dominated epoch and before structure formation the velocity dispersion of WIMPs is given by
\begin{equation}
\sigma_{\chi}\sim10^{-10}\left(\frac{100GeV}{m_{\chi}}\right)^{1/2}(1+z),
\end{equation}
where $m_{\chi}$ is the mass of the WIMP \cite{Sikivie1999}. Since the first structures are collapsing at redshifts $z\sim 60$, we obtain a primordial velocity dispersion of $\sigma_{\chi}\sim2$ cm/s. 
In the outer halo these features will persist, but in the vicinity of the sun heating by the disk stars will broaden the phase space and physical space distribution.

Since the energy increase is proportional to time (see equation (\ref{sigma})) we obtain a heating factor
\begin{equation}
\Delta\sigma=C \sqrt{t}\sim 10^{-3}\left(\frac{km}{s}\right)\sqrt{\frac{t}{Myr}},
\end{equation}
where $C$ has been determined via our simulations. For a microhalo on an average orbit we find an increase in the velocity dispersion of $\Delta\sigma\sim12$ m/s which is much larger than the primordial velocity dispersion, effectively smearing out all caustic overdensities in the vicinity of the Galactic disk.

\section*{Conclusion}

As the prospects for direct and indirect detection of WIMP dark matter improve with the development of new detectors, a renewed interest in the phase space distribution of dark matter particles has arisen. It has recently been shown that hierarchical clustering continues down to extremely small mass scales, so that most dark matter currently in the halo of our galaxy may have originated in microhalos with masses as small as $10^{-6} M_{\odot}$. Possibly dense surviving cores, tidal streams, and caustic structures might leave phase space sparsely populated, suggesting exciting new possibilities for novel signatures that differ from the traditional experimental assumption a smooth isothermal halo.
However, our results imply that tidal effects and gravitational heating effectively wipe out any such signatures for Earth based detectors. Even though we find that a significant fraction of microhalos still have a bound core today, these overdense regions are too small to be relevant for detection experiments.

The disrupted material in the tidal streams is not dense enough to affect the detection signal. On Earth, there are about $7\cdot10^8$ dark matter particles per second streaming through our bodies (assuming $m_{\chi}\sim 100$ GeV), but they originate from over $10^3$ streams coming from disrupted microhalos. The velocity dispersion in the streams is heated up through stellar interaction from initially $10^{-3}$ km/s to $10^{-2}$ km/s and therefore we expect an essentially smooth Maxwellian, with at most some spikes due to microhalos with an unusual orbital history \cite{Kuhlen2010,Vogelsberger2010}.

To summarise, our results imply that limits obtained on dark matter from detection experiments under the conservative assumption of a smooth halo with nearly Maxwellian density distribution remain valid, with a prefactor depending only on the average local dark matter density. The characteristic deviations from a Maxwellian distribution predicted from numerical simulations may be detected given sufficient detection statistics \cite{Hansen2006, Vergados2008}. Substructure cores still persist in the vicinity of the Sun to today, however the boost factor for indirect detection even for these is likely to be relatively small.

We thank Doug Potter for computational help. LMK acknowledges the hospitality of the ITP at the University of Zurich, where this work was initiated and the Pauli Center for Theoretical Studies for financial support. This research is supported by the Department of Energy Office of Science (Arizona), and the Swiss National Foundation.

\end{document}